\documentstyle[preprint,aps,a4,12pt,axodraw]{revtex}
\newcommand{\pr}{\hspace{\parindent}}
\begin{document}
\draft
\hfill\vbox{\baselineskip14pt
            \hbox{\bf KEK-TH-477}
            \hbox{\bf YUMS-96-27}
            \hbox{\bf SNUTP-96-110}
            \hbox{(\today)}}
\baselineskip20pt
\vskip 0.5cm 
\begin{center}
{\Large\bf Virtual SUSY Threshold Effects \\
and CDF large $E_{_T}$  Anomaly}
\end{center} 
\vskip 0.2cm 
\begin{center}
\large C.S. Kim$^{1,2}$\footnote{cskim@kekvax.kek.jp, 
kim@cskim.yonsei.ac.kr} ~ and ~
S. Alam$^2$\footnote{sher@theory.kek.jp} 
\end{center}
\begin{center}
1: {\it Department of Physics, Yonsei University, Seoul 120-749, Korea}\\
2: {\it Theory Group, KEK, Tsukuba, Ibaraki 305, Japan }\\
\end{center}
\vskip 0.2cm 
\begin{center} 
\large Abstract
\end{center}
\begin{center}
\begin{minipage}{14cm}
\baselineskip=18pt
\noindent

Recent CDF data  of the inclusive jet cross section shows 
threshold-like structured deviation, around
transverse momentum $E_{_T}(j)\approx 200 \sim 350 $ GeV. 
If this data is real, not just some statistical
fluctuation, is it possible to interpret the anomaly in terms of virtual
SUSY effects?
The purpose of this note is to address this
question. However, we find that virtual SUSY loop  interference effects 
[near the threshold] are too small to explain the CDF data.    
Our main conclusion seems to be on the right track if we assume that the 
recent global analysis of improved parton distributions 
by Lai {\it et al.} is correct.

\end{minipage}
\end{center}
\vfill
\baselineskip=20pt
\normalsize
\newpage
\setcounter{page}{2}

\section{Introduction}
\pr

The CDF \cite{CDF96} and D0 \cite{D096} collaborations  at Tevatron 
Collider have recently reported data for the inclusive jet cross section 
in $p\bar{p}$ collisions at $\sqrt{s} = 1800$ GeV. 
Let us recapitulate some particulars of this data 
[we concentrate mainly on the CDF data].
\begin{itemize}
\item{}The CDF measurement is based on a data sample of 19.5 pb$^{-1}$
collected in 1992-93 with the CDF detector at the Tevatron collider.
Jets were reconstructed using a cone algorithm.
\item{}Cosmic rays and accelerator loss backgrounds were removed
with cuts on event energy timing and on missing transverse energy.
The remaining backgrounds are claimed to be less 0.5\% in 
any $E_{_T}$ bin.
\item{}The measurements have been reported over a wide range 
of transverse energy, 15 GeV $\leq E_{_T} \leq$ 440 GeV, 
and around the central pseudorapidity region 
${\rm 0.1 \leq |\eta| \leq 0.7}$. The smallest distance 
probed is on the order of $10^{-19}$ m.
\item{}After accounting for uncertainties the corrected
experimental cross section, when compared to the 
Next-to-leading [NLO] QCD predictions using MRSD0$^\prime$ parton 
distribution function [PDF's], is significantly higher than
the NLO prediction for $E_{_T} > 200$ GeV. 
For $E_{_T}$ below 200 GeV the agreement between the CDF and the NLO QCD
cross section is excellent, while the D0 results are 
higher than the NLO PQCD  
predictions within the statistical uncertainties.
\item{}CDF collaboration have compared their data with
other PDF's and a model about presence of quark substructure. 
This will be discussed later.  
\end{itemize}  

There are basically two logical possibilities
for explanations of the CDF data on inclusive jet production cross sections:
(i) the parton distribution functions determined at low $Q^2$ region may not
be accurate enough to be applicable to the high $E_{_T}$ region with $E_{_T} > 
200$ GeV, or (ii) there are some new physics around the electroweak scale.
We give a brief review of  these two possibilities in the following.

Let us first consider the  first possibility.
According to the CDF collaboration
\cite{CDF96} ``the excess of data over theory at high $E_{_T}$ 
remains for CTEQ2M, CTEQ2ML, GRV94, MRSA$^\prime$, and MRSG parton 
distribution''.
The variations in QCD predictions represented a survey of 
then-available distributions. They do not represent
the uncertainties associated with data used in deriving
the PDF's. Inclusion of these new data in a global fit with 
those from other experiments may yield a consistent set of
PDF's which will accommodate the high-$E_{_T}$ excess within the scope of 
QCD. Glover {\it et al.}, \cite{Glo96} conclude that ``it is 
unlikely that the difference between the
CDF inclusive jet cross section data and standard NLO QCD
prediction can be attributed to a deficiency in our knowledge
of parton distributions''. More precisely Glover {\it et al.} \cite{Glo96}
find from their global analysis that it is impossible to fit
both the CDF data for $ E_{_T} > 200 $ GeV and Deep Inelastic 
Scattering [DIS] data for $ x > 0.3 $. However, as noted in \cite{John96}
the interpretation of large $E_{_T}$ jet cross sections inherits
uncertainties from the non-perturbative parton distribution and
fragmentation functions. It has recently been reported \cite{Lai96} that
the apparent discrepancy between CDF data and theory may be 
explained by the uncertainties resulting from the non-perturbative
parton distribution, in particular in the gluon distribution at large $x$.
These authors have also performed NLO QCD global analysis
including the CDF data and conclude that high $E_{_T}$ can be
explained in terms of a modified gluon PDF. However, we note 
that Lai et al.\cite{Lai96} use more parameters to describe 
the input gluon distribution than is usually done, whereas 
Glover et al.\cite{Glo96} assume that the gluon distribution 
has a canonical behavior [i.e. goes as $(1-x)^{n}$] at large $x$.

It is also tempting to look for a possible explanation for
the CDF data in terms of new physics. The CDF group \cite{CDF96}
have reported on a model of presence of quark 
substructure \cite{Eic83}. They have compared their data to leading
order QCD calculation including compositeness and have used MRSD0$^\prime$
parton distribution. They find a good agreement
between data and the compositeness model, for $E_{_T} > 200$ GeV,
for a substructure scale of $\Lambda_C=1.6$ TeV.

Yet another possibility is that jet measurements at hadron
colliders may be sensitive to quantum corrections due to
virtual SUSY particles \cite{Bar95,Wil96,John96,Alam96}. The
purpose of this note is to concentrate on this scenario. 
The layout and of this paper is as follows. In next 
section we discuss our calculation of the SUSY virtual 
threshold effects. The final section contains discussions 
and conclusions of our numerical results.

\section{Virtual SUSY Threshold Effects}
\pr

We consider the SUSY one loop corrections to
the process ${\rm d\sigma(p\bar{p}\rightarrow 2\;jets)}$.
As is well-known, the 2 jets production cross section in 
proton anti-proton collisions is found by weighing the expressions 
for differential cross section of the subprocess, 
$d\hat{\sigma}_{ij} \equiv d\hat{\sigma}
(ij \rightarrow 2~ {\rm  final~ partons})$, 
by the parton distribution functions, 
and integrating over the parton variables, i.e. 
\begin{equation}
d\sigma(p\bar{p}\rightarrow 2\;jets)=\sum_{i,j}\;
\int_{0}^{1}dx_1 \int_{0}^{1}dx_2
[f_{i/p}(x_1,Q^2)\;f_{j/\bar{p}}(x_2,Q^2)]
d\hat{\sigma}_{ij}(\alpha_s,\hat{s},\hat{t},\hat{u}).
\label{fe1}
\end{equation}
Here $d\hat{\sigma}_{ij}$ represents the subprocess cross section
at c.m. energy square of $\hat{s}=x_1 x_2 s$, where $\sqrt{s}$ is the
c.m. energy of the $p\bar{p}$ system. 
It is well-known that SUSY particles [gluinos, squarks] decrease
or slow down the rate of fall of $\alpha_s(\mu)$ for large scale $\mu$.
``Large'' means far above the threshold. At the one loop
level the evolution equation for $\alpha_s(\mu)$ can be written as
\begin{equation}
\frac{d}{d\ln \mu}\alpha_{s}(\mu)=-\frac{b_{3}}{2\pi}
\alpha_{s}^{2}(\mu)~.
\end{equation}
In the SM, $b_3$ is given by
\begin{equation}
b_3=11-\frac{2}{3}n_f~,
\end{equation}
whereas in MSSM model one has
\begin{equation}
b_3=11-\frac{2}{3}n_f-2-\frac{1}{3}\tilde{n}_f~,
\end{equation} 
where $n_f$ [$\tilde{n}_f$] is the number of quark [squark] flavors that
are active. The contribution `$-2$' is the gluino contribution [it is assumed
that the gluino is active in this case]. 

With this in mind, we can see that the SUSY corrections to Eq.~(\ref{fe1})
can be broken into three parts, namely  (a): SUSY corrections to 
the PDF, (b): SUSY corrections to the running of $\alpha_s$,  
(c): parton-level SUSY loop corrections to $d\hat{\sigma}_{ij}$,
excluding the corrections already included in (b) [i.e. the running 
of $\alpha_s$]. 
And finally we have to (d): combine those three part, as in Eq. (\ref{fe1}),
by convoluting with PDF to get hadron-level SUSY interference effects. 
In this note we consider all these four parts [i.e. (a),(b),(c),(d)].
Previous works on SUSY corrections \cite{Bar95,Wil96,John96} 
have considered only part of them:
In Ref.~\cite{Bar95} only the issue of the
effect of high-mass thresholds due to gluinos, squarks and other
new heavy quarks on the evolution of 
$\alpha_s$ was considered [i.e. (b)].
The corrections to $\alpha_s$ were found to be appreciable, this
in turn means a significant increase in the transverse momentum
dependence of jet production at the Tevatron. However, as noted
in Ref.~\cite{Wil96}, these  authors \cite{Bar95} do not include
the effect of $q \tilde{q} \tilde{g}$ Yukawa interactions, 
and hence one cannot take their
estimates for the superpartners of ordinary matter as final. 
In Ref.~\cite{Wil96} the effect of Yukawa couplings was included,
and found that the CDF data cannot
be explained by a mass threshold effect in the MSSM, as the
calculated result is not only small but of the wrong sign, considered 
at the parton level below the threshold energy scale [i.e. (b),(c)].
In a similar but more detailed analysis, the authors of
Ref.~\cite{John96}, working in the context of MSSM, consider
at around the threshold energy scale the virtual one-loop corrections 
to the parton-level subprocesses
$q\bar{q}\rightarrow q\bar{q}$, 
$q\bar{q}\rightarrow q^\prime \bar q^\prime$,
$qq^\prime\rightarrow qq^\prime$, 
$q\bar{q}\rightarrow gg$ and $qg\rightarrow qg$, 
which are expected to dominate the large $E_{_T}$ cross sections at
the Tevatron energy [i.e. (b),(c)].

The purpose of this note is to give our results
of incorporating the one-loop radiative corrections into
the running of $\alpha_s$, the dressing-up of the parton
distribution functions, and finally convoluting the relevant
subprocess cross sections 
with the SUSY dressed-up PDF's [i.e. (a),(b),(c),(d)]. 
In the hadron colliders, like the 
Tevatron, what is measured is $p\bar{p}$ cross section, 
and not the individual subprocesses cross sections.
So in order to determine the effect of
subprocesses on the $E_{_T}$ cross section one must perform
a convolution of the cross section of each subprocess
with the corresponding PDF's [i.e. (d)]. 
We find that in the process 
of convolution with the PDF's, the ``dips and peaks'' in the various 
subprocesses \cite{John96} are much reduced.
   
We note that, also as pointed out in \cite{John96}, one 
should take into account sparticles effects on the parton
structure functions at energies {\bf sufficiently far above}
the threshold, and can ignore this effect around the threshold
region [i.e. (a)]. We test the validity of this statement and find it
to be true from our numerical work, as will be shown in Fig. 2. 
We have considered the combined evolution equations 
for $\alpha_s$, $q_v$ 
\begin{equation}
\frac{d}{d\ln Q^2}\alpha_{s}(Q^2)=-\frac{b_{3}}{4\pi}
\alpha_{s}^{2}(Q^2)~,
\end{equation}
\begin{equation}
\frac{d}{d\ln Q^2}q^{NS}(x,Q^2)=\frac{\alpha_s(Q^2)}{2\pi}
P_{0}^{NS}\otimes q^{NS}(x,Q^2)~.
\end{equation}
The second equation becomes, with the definition of
$p^{NS}=xq^{NS}$,
\begin{eqnarray}
\frac{d}{d\ln Q^2}[p^{NS}(x,Q^2)]&=&\frac{2\alpha_s(Q^2)}{3\pi}
\int_{x}^{1}\frac{dz}{1-z}[(1+z^2)[p^{NS}(x/z,Q^2)]
-2p^{NS}(Q^2,x)]\nonumber\\
& & +\frac{\alpha_s(Q^2)}{\pi}[1+\frac{4}{3}\ln(1-x)]
p^{NS}(x,Q^2)~.
\end{eqnarray}
The increase of 
$\alpha_{s}(Q^{2})$ results in the decrease of the  PDF compared to the 
SM when we make the  evolution. The qualitative reason is rather simple: 
stronger $\alpha_s (Q^{2})$ as $Q^2$ increases imply that the gluon radiation
from the initial quarks are enhanced, and the PDF evolution yields the larger
gluon or sea-quark densities at the smaller  $x$ region, and therefore the
valence quark distribution $q_{v}(x,Q^{2})$ decreases at large $x$,
as $\alpha_{s}(Q^{2})$ increases, and vice versa.  
We recall that for large $x$ the valence contribution dominates
(e.g. $q_v \cdot q_v : q_v \cdot g : g \cdot g \sim 0.65 : 0.3 : 0.05$
at $x \sim 0.3$ at Tevatron energies),
we ignored the SUSY evolution of sea-quark 
or gluon for large $E_{_T}$ of CDF.
It turns out, as shown in Fig. 2 and claimed in Ref. \cite{John96},
the PDF with sparticle effects at around the threshold region deviates
much less than 0.1\% from its SM predictions,  which justifies 
our assumption. And we can even totally ignore the SUSY corrections to PDF
[i.e. (a)] 
all together for investigations below or around the threshold region.

\section{Numerical Results and Conclusions}
\pr

For our numerical calculation, we implement various lower bounds 
on squarks and gluinos depending on parameters in the MSSM.  For example, 
D0 group \cite{D0SUSY} searched for the events with large missing $E_{_T}$ 
with three or more jets, observed no such events above the level 
expected in the SM.
This puts some limits on the squark and gluino masses assuming 
the short-lived gluinos:
\begin{eqnarray}
m_{\tilde{g}} & > &  144 ~{\rm GeV~~~for~~~} m_{\tilde{q}} = \infty~,
\\
{\rm or}~~~~m_{\tilde{g}} & = & m_{\tilde{q}} > 212~{\rm GeV}~. 
\label{eq8}
\end{eqnarray}
CDF group is currently analyzing their data, with their
preliminary data being similar to the D0 results  with slight increase in
sparticles' mass bound. 
In the subsequent numerical analysis, we choose three sets of 
$(m_{\tilde{g}}, m_{\tilde{q}})$, which we shall refer to as 
Case I, II and III, respectively,
\begin{equation}
(m_{\tilde{g}}, m_{\tilde{q}}) =  (220 ~ {\rm GeV}, 220~ {\rm GeV}),~~
(150~ {\rm GeV}, \infty),~~ (150 ~ {\rm GeV}, 150~ {\rm GeV})~.
\label{eq9}
\end{equation}
The Case III is only of academic interest if the limit given in 
Eq. (\ref{eq8}) is valid in reality.

We turn now to our numerical result. The solid curve in Fig.~1
represents the one-loop evolution of $\alpha_s(\mu)$ versus the 
renormalization scale $\mu$ for the SM, assuming a top-quark mass
of 175 GeV, and $\Lambda_{QCD}=0.2$ GeV. The long-dashed curve 
represents the results for the MSSM with the five squarks and gluinos 
being degenerate with a common mass of 220 GeV [we shall refer to
this as Case I]. As can be seen from Fig.~1, Case I starts to deviate 
from the SM results at $\mu=440$ GeV, as expected. The deviation
in the value of $\alpha_s$ away from the SM is maximum [for our range
of $\mu$] at $\mu=700$ GeV and roughly on the order of $5\%$.  
This value is to be compared to the $12\%$ reported in \cite{Bar95}.
The dotted curve in Fig.~1 describes our result for the MSSM with
gluino mass of 150 GeV, and the squarks are assumed to be decoupled,
we shall refer to this as Case II. Case II starts to be different
from the SM at $\mu=300$ GeV. The maximum deviation for Case II
occurs at around 700 GeV, where it also approaches the curve for
Case I.

The deviation from the SM of the one-loop evolution for 
the parton distribution function $[{\rm (\frac{MSSM-SM}{SM})u_v(Q)}]$
versus the factorization scale Q for Cases I and II are shown in 
Fig.~2. Case I deviates from the SM at $Q=440$ GeV attaining
a maximum change of $-0.05\%$ at 700 GeV. The magnitude of 
deviation is much smaller compared to the results of Case I 
displayed in Fig.~1. The sign of corrections is also opposite
to those for Case I in Fig.~1. This indicates as that $\alpha_s$
increases [as for {\it e.g.} it does for MSSM] $u_v(Q)$ decreases. This
is natural to expect since for larger $\alpha_s$ there is more
probability of gluon radiation, which in turn implies larger
gluon and sea densities at small x, which eat away at the $u_v$
contribution thus resulting in a reduction in $u_v$.
Turning to Case II, we see from Fig.~2 that it deviates
from the SM at $Q=300$ GeV, reaching a maximum value of only $-0.1\%$ at
700 GeV.      
 
Fig.~3 exhibits our results for $[{\rm (\frac{MSSM-SM}{SM})
\frac{d\sigma}{dp_{_{T_j}}}}]$ versus ${\rm p_{_{T_{j}}}}$ for
both Cases I and III [Case III refers the MSSM with the five squarks 
and gluinos being degenerate with a common mass of 150 GeV, {\it i.e.} 
the last set in Eq. (\ref{eq9})]. 
A rapidity cut of 0.7 is used [$|\eta|<0.7$].
A maximum difference of $-1\%$ from the SM is found for the both 
Cases. The question that naturally arises is, that why the  
``dips and peaks'' [which are at the level of 5\%--6\% \cite{John96} ]
at the parton level, are reduced. 
There are two factors contributing to the reduction in the
``dips and peaks'' in the process of going from the parton to the 
process level. Part of this reduction comes from the process of
convolution of the various subprocesses with PDF's [as already 
remarked to in the previous section].  
The other piece of this reduction in ``dips and peaks''
by going from subprocess to process level is due to t-channel
``dilution'' effect. This can be shown by simply not including
the t-channel subprocesses' contribution. When this is done [see
Fig.~4a and 4b] reduction in the ``peaks and dips ''is not so large.

In Fig.~4(a) we exhibit our results for $[{\rm (\frac{MSSM-SM}{SM})
\frac{d\sigma}{dp_{_{T_j}}}}]$ versus ${\rm p_{_{T_{j}}}}$ for
both Cases I and III, but this time only including the subprocesses
$q\bar{q}\longrightarrow q'\bar{q}'$ and $q\bar{q} \longrightarrow gg$.
These two are the subprocesses with the prominent threshold structures from
vacuum polarization interferences through $s$-channel exchange diagrams. 
A rapidity cut of 0.7 is used [$|\eta|<0.7$].
The deviation from the SM for Case I varies between 2\%  to $-4\%$,
whereas as in Case II the variation is similar size to Case I at 
different energy scales related to the SUSY threshold effects. 
In Fig.~4(b) we exhibit our results for $[{\rm (\frac{MSSM-SM}{SM})
\frac{d\sigma}{dM_{jj}}}]$ versus $M_{jj}$ for both the Cases I and III,
including the subprocesses
$q\bar{q}\longrightarrow q'\bar{q}'$ and $q\bar{q} \longrightarrow gg$.
The percentage variation is almost the same as in Fig.~4(a).

Near the virtual (or direct) SUSY threshold, 
the Coulomb interactions between the gluinos [or squarks] with a 
very small velocity $v = \sqrt{1 - 4 m_{\tilde g}^2 / Q^2}$ make the QCD 
effects depending on the parameter $\alpha_{s} / v$, 
rather than on $\alpha_s$.
The result of summing up the Coulomb exchanges  reduces to multiplying
the Coulomb factor $\alpha_{s} /v$~ \cite{Coulomb}, 
which gives a wide resonance structure 
around $Q \approx 2 m_{\tilde g}$ for the short-lived gluino pairs.  
If the life time of gluino is long enough, then narrow bound states of 
gluino pair may appear near the threshold region. 
Since the  properties on gluino pair bound states
are not known and largely model dependent, we can guess only qualitative 
nature of the threshold region, $Q \approx 2 m_{\tilde g}$ 
(or $Q \approx 2 m_{\tilde q}$), as shown in Fig.~3 and 4.  

In summary, in the MSSM the $E_{_T}$ distributions does not
differ very much from those of the SM except for the possible 
threshold effects ($\sim 1\%$) through loop corrections.  
In actual experiments, the jet resolution
will in general smear out any narrow resonance structures (which may be the
case for the long-lived gluinos), leading to broad resonance structure,
and therefore it looks impossible to detect SUSY particles through this kind
of indirect virtual threshold effect.  
As is previously explained, it has been reported \cite{Lai96} that
the apparent discrepancy between CDF data and theory may be 
explained by the uncertainties resulting from the non-perturbative
parton distribution, in particular in the gluon distribution.
Our main conclusion seems to be on the right track in view
of this  global analysis of parton distribution function.

\section*{Acknowledgements}

The authors would like to thank E.L. Berger, M. Drees,  
P. Ko, Jake Lee and J.W. Qui for useful suggestions and helpful comments.
The work of CSK was supported 
in part by the KOSEF, 
Project No. 951-0207-008-2, in part by CTP of SNU, 
in part by the BSRI Program, Project No. BSRI-97-2425, and
in part by the COE Fellowship from Japanese Ministry of 
Education, Science and Culture.
The work of SA is supported by COE fellowship of the Japanese
Ministry of Education, Science and Culture.




\section*{Figures}

\begin{itemize}

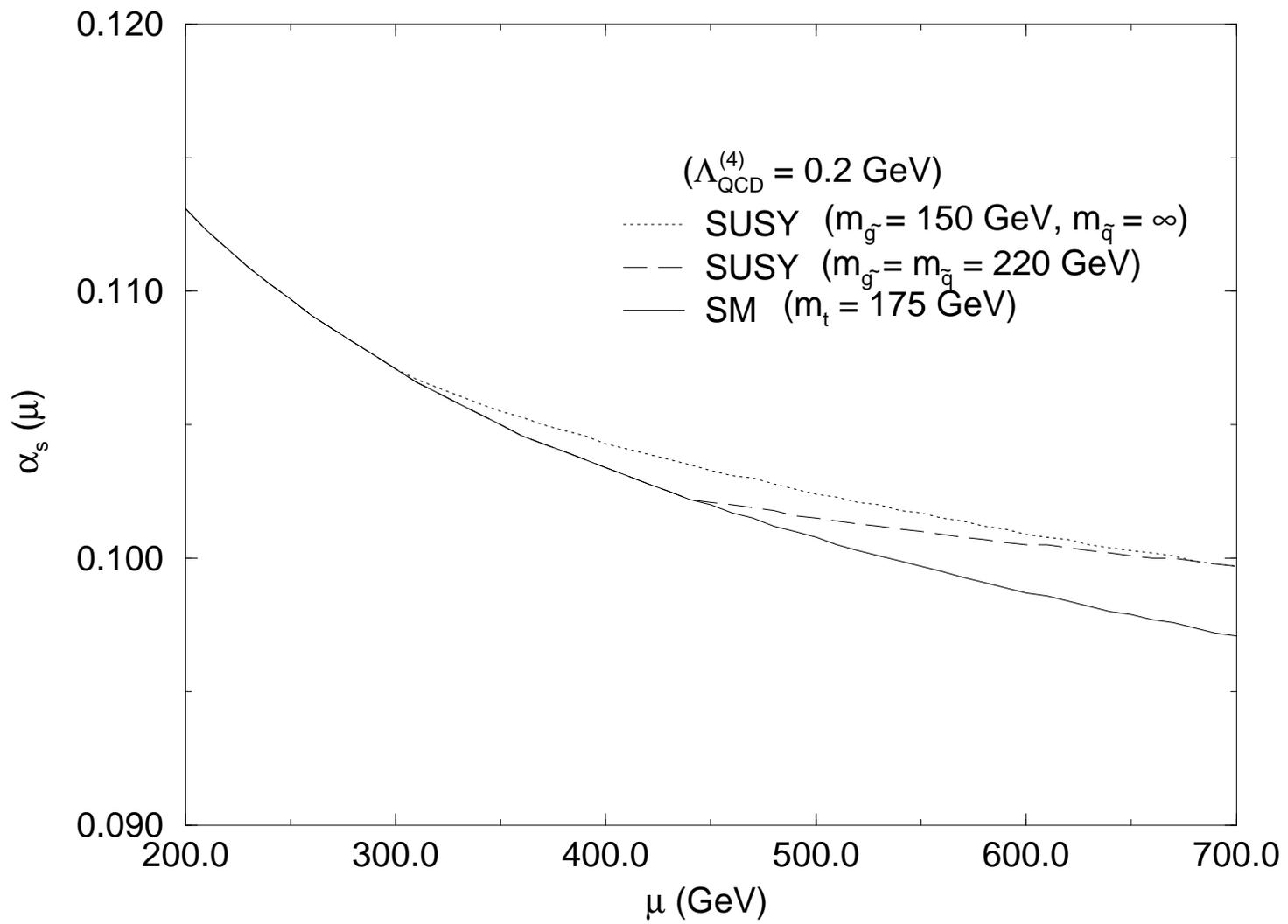
\item[{\bf Fig.~1}] The one-loop evolution of $\alpha_s(\mu)$
versus $\mu$ is given for SM [solid line] and SUSY cases.

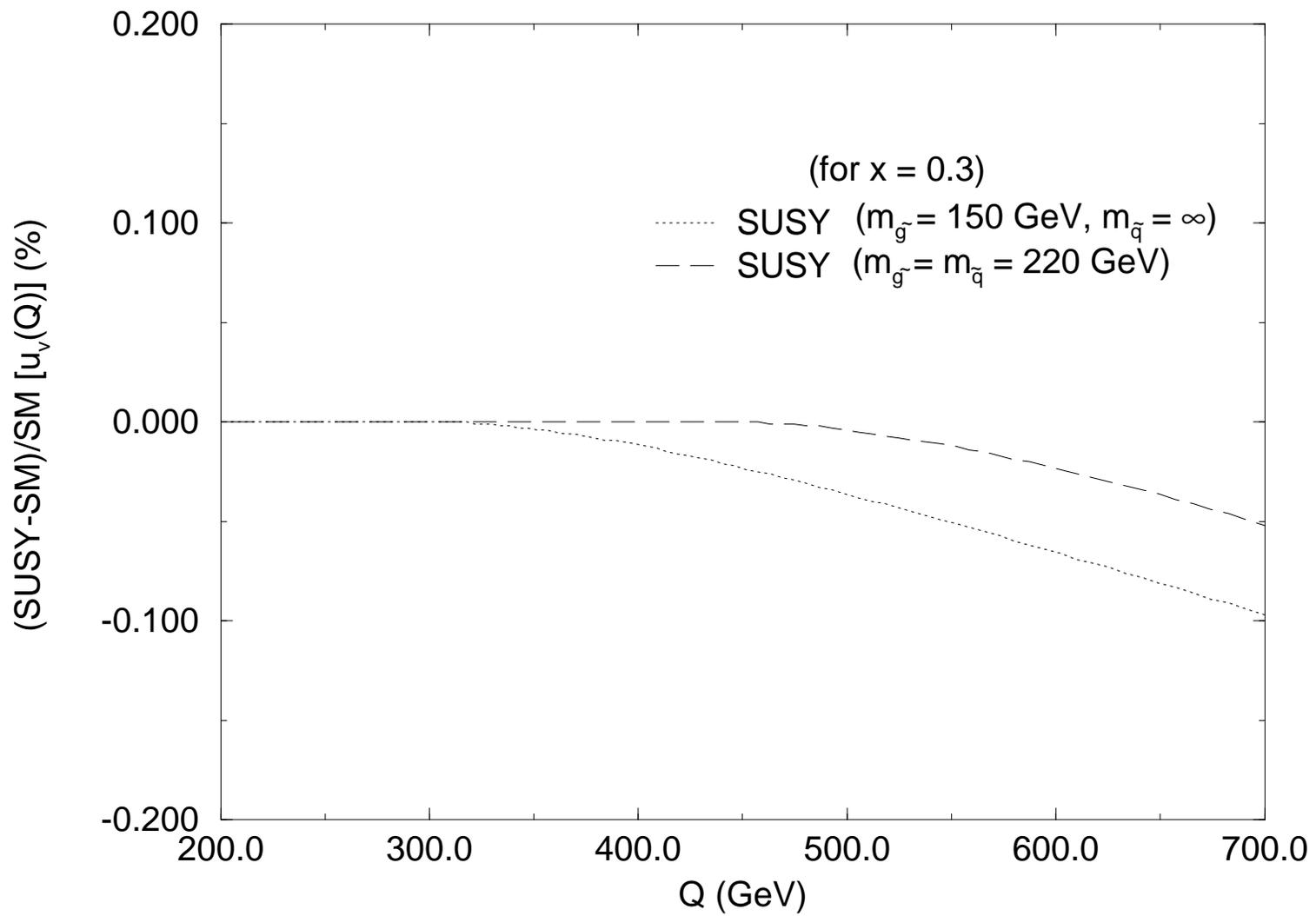
\item[{\bf Fig.~2}] Percentage deviation from the SM due to SUSY
contribution of PDF of the valence quark versus 
factorization scale Q for typical momentum fraction $x=0.3$.  

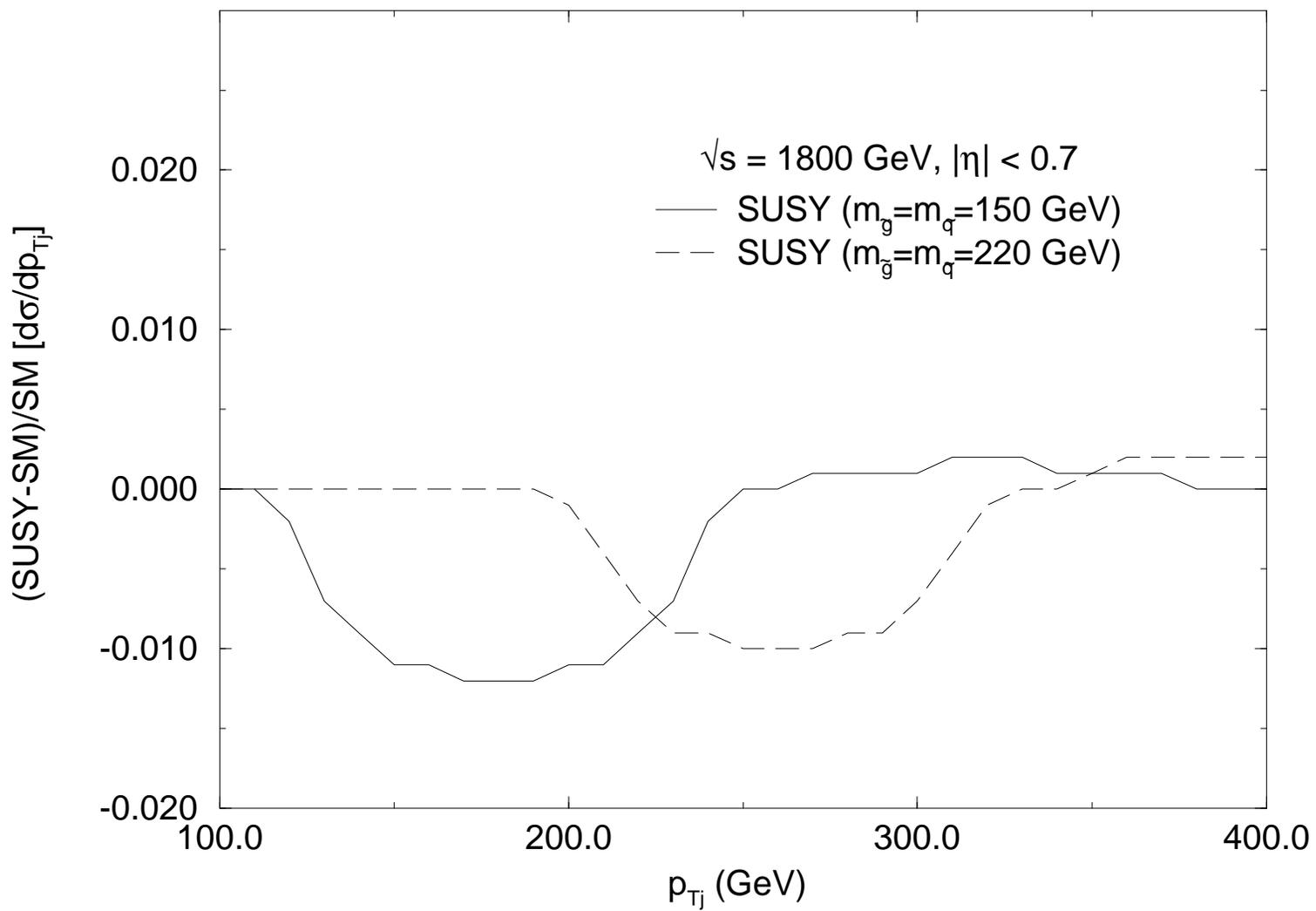
\item[{\bf Fig.~3}] Deviation from the SM due to SUSY
contribution to $\frac{d\sigma}{dp_{_{T_j}}}$ versus $p_{_{T_j}}$.
  
\item[{\bf Fig.~4a}] Deviation from the SM due to SUSY
contribution to $\frac{d\sigma}{dp_{_{T_j}}}$ versus $p_{_{T_j}}$, when only 
the subprocesses $q\bar{q}\rightarrow q^{\prime}\bar{q}^{\prime}$,
and $q\bar{q}\rightarrow g g $ are included.
  
\item[{\bf Fig.~4b}] Deviation from the SM due to SUSY
contribution to $\frac{d\sigma}{dM_{jj}}$ versus $M_{jj}$, when only 
the subprocesses $q\bar{q}\rightarrow q^{\prime}\bar{q}^{\prime}$,
and $q\bar{q}\rightarrow g g $ are included. 

\end{itemize}

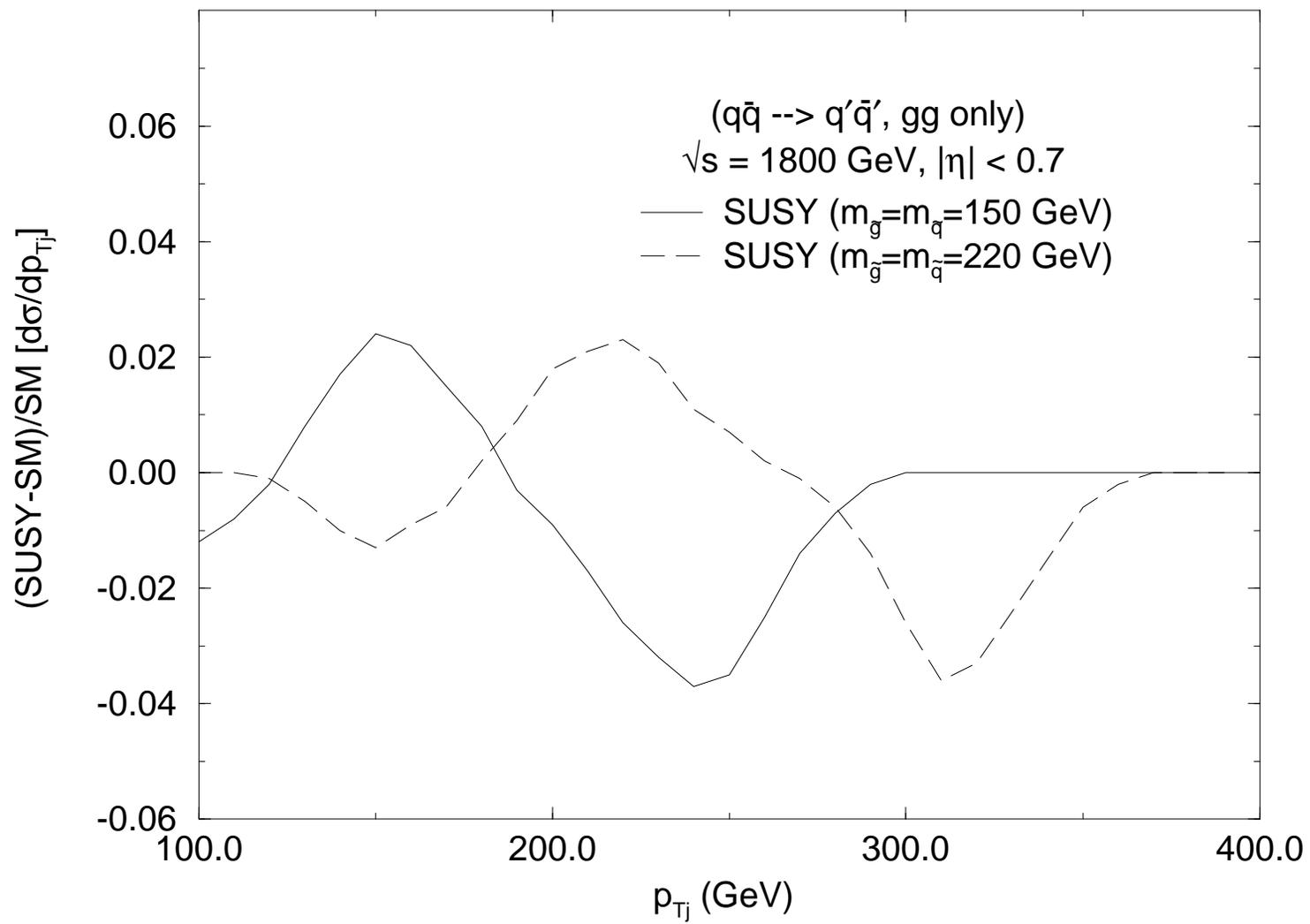
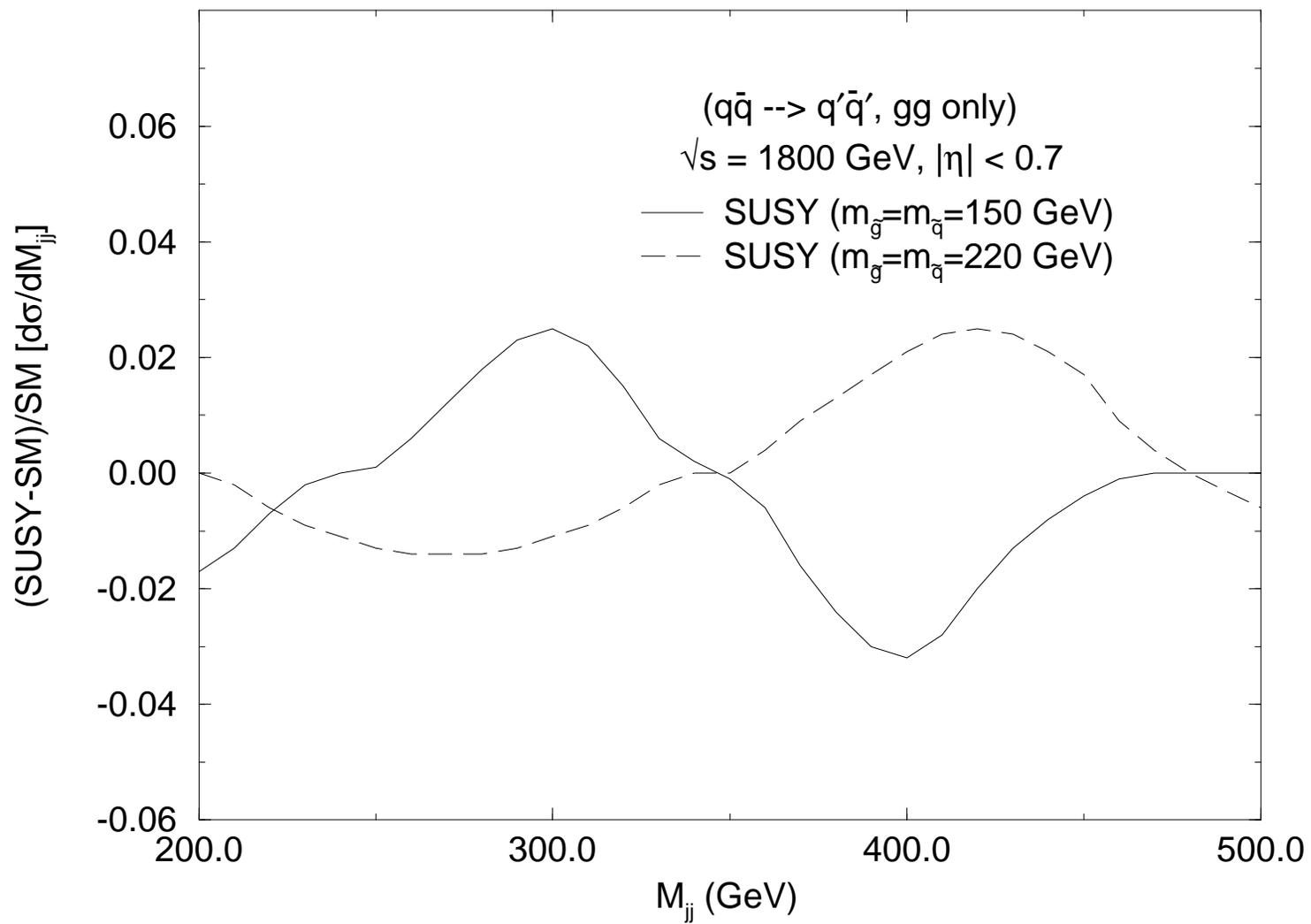
\end{document}